# Absolute frequency measurement of an $SF_6$ two-photon line using a femtosecond optical comb and sum-frequency generation.

Anne Amy-Klein, Andrei Goncharov♣, Mickaël Guinet, Christophe Daussy, Olivier Lopez,

Alexander Shelkovnikov♦, Christian Chardonnet

Laboratoire de Physique des Lasers, UMR 7538 CNRS, Université Paris 13, 99 av. J.-B.

Clément, 93430 Villetaneuse, France

Abstract : We demonstrate a new simple technique to measure IR frequencies near 30 THz using a femtosecond (fs) laser optical comb and sum-frequency generation. The optical frequency is directly compared to the distance between two modes of the fs laser, and the resulting beat note is used to control this distance which depends only on the repetition rate fr of the fs laser. The absolute frequency of a CO2 laser stabilized onto an SF6 two-photon line has been measured for the first time. This line is an attractive alternative to the usual saturated absorption OsO4 resonances used for the stabilization of CO2 lasers. First results demonstrate a fractional Allan deviation of 3×10-14 at 1 s.

*OCIS codes :*

---

♣ Permanent address : The Institute of Laser Physics, Siberian Branch of the Russian Academy of Science, Pr. Lavrentyeva 13/3, 630090 Novosibirsk, Russia.
♦ Permanent address : P.N.Lebedev Physical Institute, Leninsky Prospect, 53, Moscow, 117924, Russia.

The frequency comb provided by femtosecond (fs) lasers is now widely used for optical frequency measurements [1]. The most common scheme involves the comparison of the optical frequency to be measured with the nearest mode of the comb, and needs control of the absolute frequency of the comb. The frequency $f_p$ of the $p^{th}$ mode of the comb depends on two radiofrequencies : $f_p = p\, f_r + f_0$ where $f_r$ and $f_0$ are respectively the repetition rate and the comb frequency offset of the fs laser, and p is an integer around $10^6$. The repetition rate is easily detected with a fast photodiode, while the self referencing technique is commonly used for the detection of $f_0$, which needs a broadening of the frequency comb to more than one octave [2]. An alternative method is to use a second laser as an optical reference [3]. For the measurement of infrared frequencies, another scheme can be implemented which is based on sum-frequency generation (SFG) in a nonlinear crystal. The absolute optical frequency is converted into a frequency difference : it is compared to the difference between two modes of the comb, that is to a very high harmonic of the repetition rate. This scheme is independent on the comb offset $f_0$ and does not require any broadening of the comb for infrared frequencies around 30 THz. It was first demonstrated at 10 μm with a $CO_2/OsO_4$ stabilized laser[4], then with an HeNe/$CH_4$ stabilized laser at 3.39 μm[5, 6] and with an acetylene-stabilized laser at 1.5 μm [7]. An alternative scheme using the generation of an offset-free difference-frequency comb in a nonlinear crystal could also be used[8].

Here we demonstrate a simplified version of a molecular optical clock operating with a $CO_2$ stabilized laser at 28 THz. Our previous scheme[4] combined SFG and the use of two laser diodes as intermediate oscillators. A laser diode at 852 nm was phase locked to a fs mode. The sum of this diode and the $CO_2$ laser frequency was generated in a crystal of $AgGaS_2$ and a second laser diode, at 788 nm, was phase locked to the sum. Finally a second fs mode was phase locked to the diode at 788 nm by feeding back to the fs cavity length. As a result, the $CO_2$ laser controlled the separation between two modes of the comb. With this scheme 2 of the 3 phase-lock loops are in fact sensitive to the offset frequency $f_0$, although it was not necessary to detect or to control $f_0$.

Our new scheme uses only one phase-lock loop and no laser diodes. The basic principle is to perform the sum-frequency generation (SFG) of the fs laser comb and the $CO_2$ laser in a nonlinear crystal. The resulting frequency comb can be expressed as $f_q^{SFG} = qf_r + f_0 + f(CO_2)$. This SFG comb overlaps the high frequency part of the initial comb, and the beat notes $f_q^{SFG} - f_p = f(CO_2) - (p-q)f_r$ are obtained, which are insensitive to $f_0$. A large number of (q, p) pairs gives the low frequency beat note $f(CO_2) - mf_r$ which is used to phase-lock the $m^{th}$ harmonic of the repetition rate to the $CO_2$ laser frequency, thus building a molecular clock.

A schematic view of the experimental apparatus is given in Figure 1. The fs Ti:Sa laser (GigaJet from Gigaoptics) emits 650 mW with a repetition rate around 1 GHz. Its spectrum spans 30 nm (or 25 THz) (FWHM) around 800 nm. About 70 mW of fs laser, and 100 mW of $CO_2$ laser are focused in a 15 mm long crystal of $AgGaS_2$ for type I SFG. The measured efficiency is around 0.5 mW/W$^2$, and phase-matching bandwidth is about 1 nm (or 500 GHz). The sum comb is then combined with the initial fs comb with an adjustable phase delay in order to compensate the crystal dispersion. A few hundreds of mode pairs give rise to a beat note between the $CO_2$ laser frequency and the 28410$^{th}$ harmonic of the repetition rate $f_r$, with a signal to noise ratio (SNR) of 20dB in a bandwidth of 100 kHz. The beat note is 20 dB weaker than with the former scheme, but could be increased with an optimization of the SFG (use of GaSe instead of $AgGaS_2$, shorter crystal) and a better adaptation of the temporal length of the initial and SFG combs. The signal is amplified by 40 dB using a tracking oscillator working around 200 MHz (bandwidth of tracking loop is 1 MHz), and is finally used to lock the fs laser repetition rate to the $CO_2$ laser frequency.

To complete the frequency measurement procedure, the repetition rate is detected with a fast photodiode and counted against a local oscillator at 1 GHz. This latter is phase-locked to a reference signal transmitted via a 43-km long optical fiber from the SYRTE laboratory, located in Paris [4]. This laboratory developed a high stability oscillator, which is based on a combination of a cryogenic Sapphire oscillator (CSO), an H-Maser and a set of low noise microwave synthesizers [9]. Its frequency is steered by the H-Maser in the long-term, and monitored by the Cs atomic fountain for accuracy[10]. This signal shows a frequency stability slightly below $10^{-14} \times \tau^{-1}$ in the range 1-10 s, and $1 \times 10^{-15}$ from 10 to $10^5$ s, see Allan deviation as dashed line in Fig. 2. The transfer through the optical link degrades this stability by less than one order of magnitude, Allan deviation as ▲ in Fig. 2, while the phase noise introduced by the link can be efficiently suppressed with an active correction [11]. For the measurement reported here the transfer was done at 1 GHz and the optical link was free of any active control.

This set-up is used to measure the R(47) $A_2$ two-photon resonance of the $2\nu_3$ band of $SF_6$ [12, 13]. This measurement forms part of our program to develop a new robust and efficient stabilization scheme for $CO_2$ lasers. It is a first step towards the improvement of their spectral purity for very high resolution spectroscopy, which is essential for high sensitivy tests of fundamental physics with molecules. This two-photon line is an attractive alternative to the usual saturated absorption $OsO_4$ resonances used for the stabilization of $CO_2$ lasers [14, 15]. The $SF_6$ gas has the advantage of being less reactive than $OsO_4$. Further, molecules of all velocities contribute to this two-photon resonance, and the excitation probability is quite high due to the very small detuning of 16 MHz of the intermediate level of the two-photon transition.

We used our usual stabilization scheme (see Fig. 1) which includes a Fabry-Perot cavity (FPC) containing the molecular gas and an electro optic modulator (EOM) to perform the various frequency shifts and modulations for the locking loops [15]. FPC gives a gain factor proportional to the finesse (about 100 in our set-up) on the SNR. Figure 3 displays the third harmonic of the 2-photon resonance, used for the $CO_2$ laser frequency stabilization. The experimental parameters are : 50 mW inside the FPC, pressure $3\times10^{-2}$ Pa, 20 kHz HWHM for the two-photon line, SNR of 1000 in 1 kHz bandwidth.

First results demonstrate a fractional Allan deviation of $3\times10^{-14}$ at 1 s, which reaches a minimum value of $6\times10^{-15}$ at 30 s and then increases proportionally to $\tau$ due to the linear drift of the laser frequency (Figure 2). At short time, it is probably limited by the noise of the optical link used for the transmission of the primary reference, (▲) in Figure 2. This Allan deviation is slightly better than with a saturated absorption line of $OsO_4$ as a reference. Main progress is the daily stability of this Allan deviation, because $OsO_4$ was quite reactive and its error signal degraded after a few hours. The short-term stability might be further improved by using a high frequency acousto optic modulator to reach the line instead of the EOM which has a very low efficiency. This will give the possibility of increasing the laser power for the 2-photon absorption.

The absolute frequency measurement of the R(47) two-photon resonance of $SF_6$ was performed by repeating the $CO_2/SF_6$ frequency measurement during a period of 8 months, as shown on Fig. 4. First series of measurements (in 2004) used the previous set-up with the two laser diodes, when the second series in 2005 used the present scheme. The mean values for both series coincides to better than 2 Hz. The mean value of all the measurements is 28 412 881 552 402 ± 44 Hz, where the uncertainty is the 1-σ deviation of the data. This is consistent with the data $\nu_{ref}$ = 28 412 881,6 ± 1,0 MHz calculated from references [12, 16]. The uncertainty of $1.5 \times 10^{-12}$, limited by the reproducibility of the $CO_2/SF_6$ stabilization, is the same order of magnitude as with $OsO_4$ [4; 11]. We estimate that the day to day reproducibility is mainly limited by the instability of the optical background . In particular, due to the low efficiency of the EOM, the carrier residual is still stronger than the sidebands at the input of the FPC and higher order modes can enter the cavity with an efficiency with depends on the optical alignment and its stability. This could affect the baseline of the reference signal. In addition, frequency shifts could be related to diaphragm effects in the FPC [17].

A simple set-up has been demonstrated for the absolute measurement of a molecular resonance around 30 THz. It includes a fs laser for which only the repetition rate must be controlled. The whole measurement apparatus is very robust and can be used continuously for several hours. As a first application, a 2-photon line of $SF_6$ has been measured with a stability of $3 \times 10^{-14}$ at 1 s limited by the $CO_2/SF_6$ stability. Further applications include the characterization of the metrological performance of a 2-photon resonance detected in a Ramsey scheme [18] and the test of the possible temporal variation of fundamental constants using such a 2-photon molecular line [19].

The authors would like to acknowledge SYRTE for providing the reference signal from the primary standard for the absolute frequency measurements, through the optical link between the two laboratories.

**Figure caption**

Figure 1 : Experimental set-up. AOM : acousto optic modulator, EOM : electro optic modulator, sb : sideband, FPC : Fabry-Perot cavity, SFG : sum-frequency generation, osc. : oscillator, CSO : cryogenic sapphire oscillator.

Figure 2 : Fractional $CO_2/SF_6$ frequency stability as given by the Allan deviation (■) of the repetition rate $f_r$ of the fs comb calculated from a series of 1-s gate measurements. For comparison is shown typical Allan deviation of the optical link (▲) and of the RF reference (dotted line).[11]

Figure 3 : R(47) two-photon line of $SF_6$ (third harmonic detection), detected in transmission of a FPC and used for frequency stabilization. Power inside the FPC : 50 mW, pressure $3\times10^{-2}$ Pa, 1 ms per point.

Figure 4 : Frequency measurement of the $CO_2/SF_6$ stabilized laser; the dotted line separates the measurements performed with the old scheme from those performed with the present scheme.

Figure 1

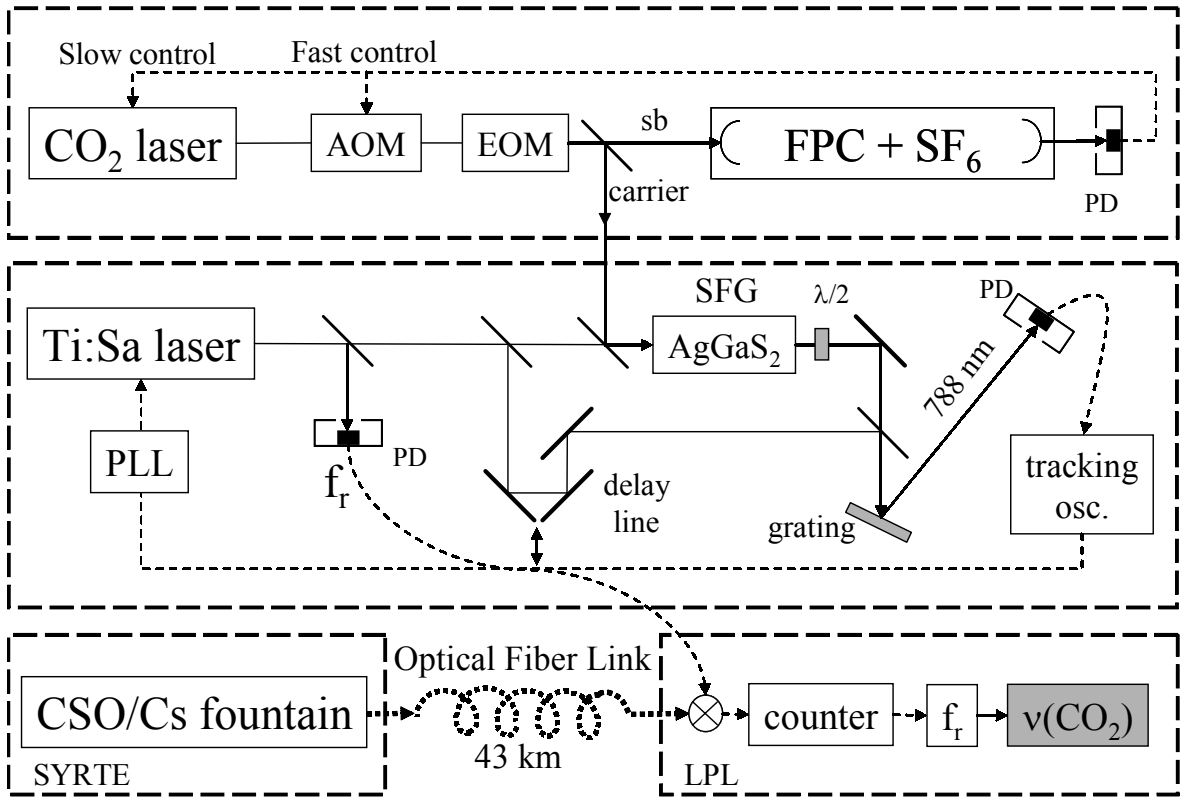

Figure 2

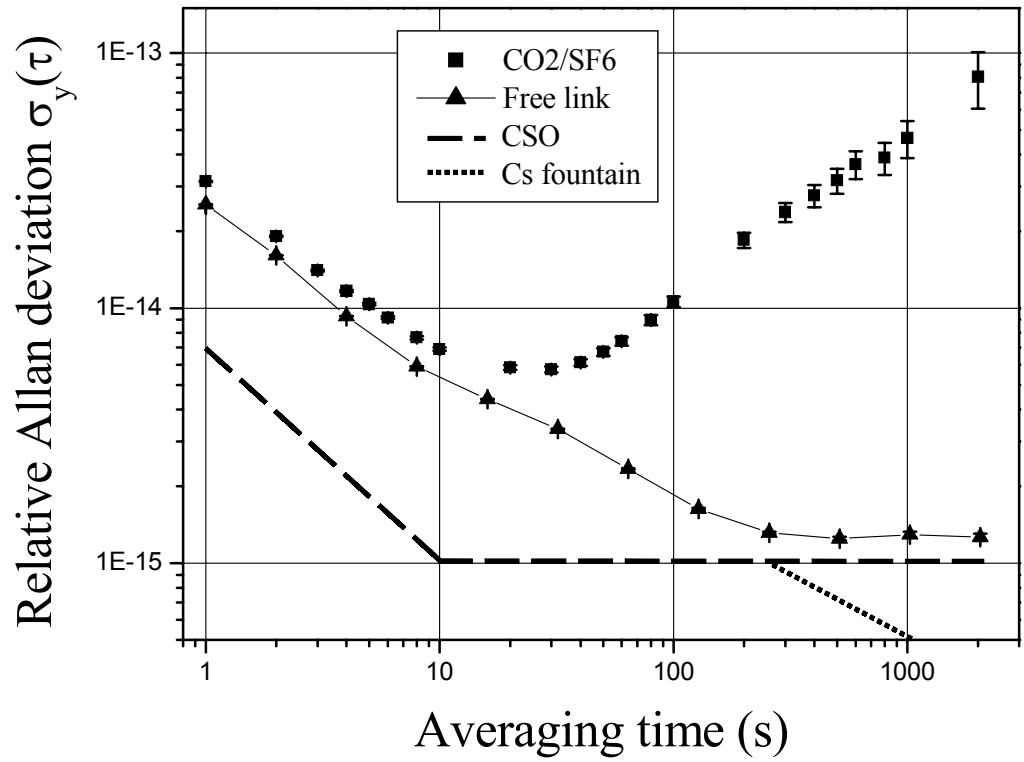

Figure 3

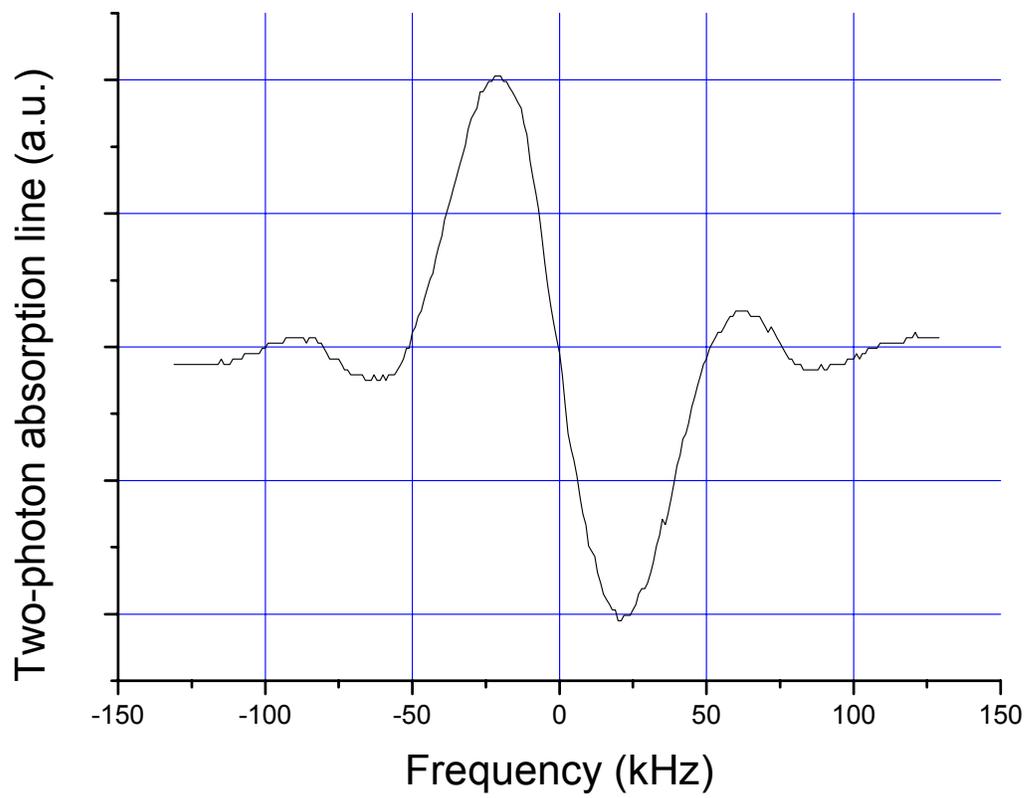

Figure 4

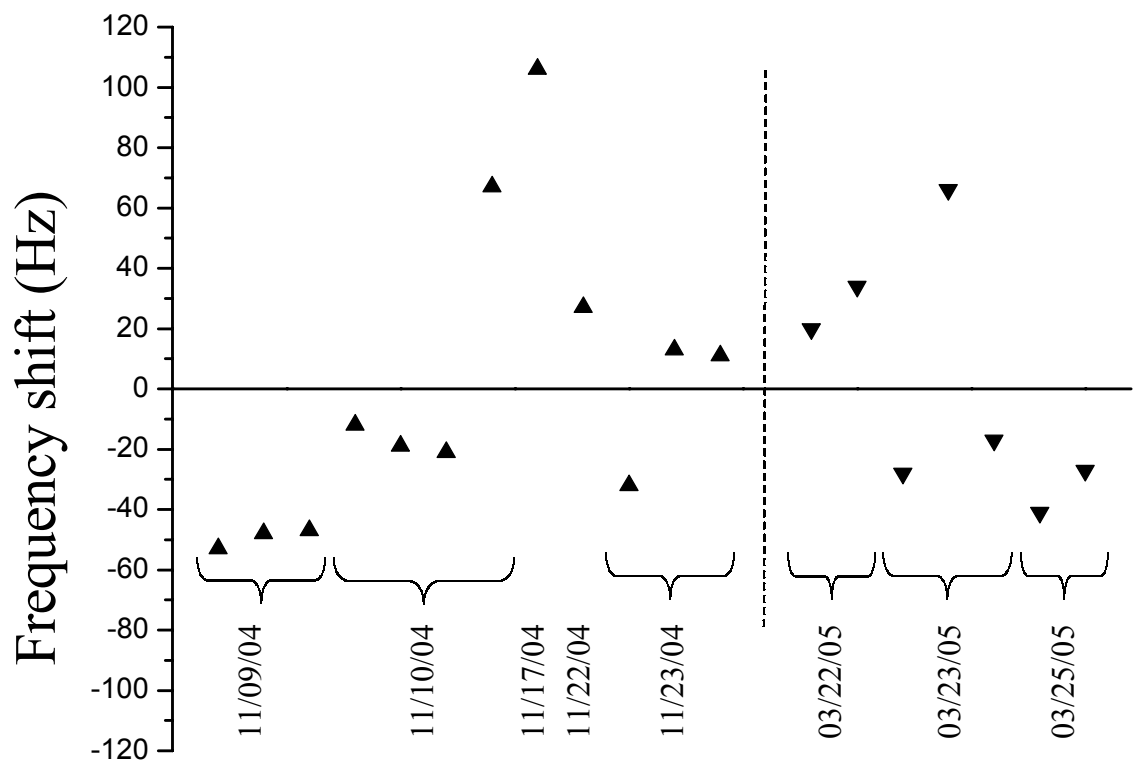